\documentclass[%
 aip,
 amsmath,amssymb,
 reprint,%
]{revtex4-1}

\usepackage{graphicx}% Include figure files
\usepackage{dcolumn}% Align table columns on decimal point
\usepackage{bm}% bold math
\usepackage[utf8]{inputenc}
\usepackage[T1]{fontenc}
\usepackage{mathptmx}
\usepackage{etoolbox}

\makeatletter
\def\@email#1#2{%
 \endgroup
 \patchcmd{\titleblock@produce}
  {\frontmatter@RRAPformat}
  {\frontmatter@RRAPformat{\produce@RRAP{*#1\href{mailto:#2}{#2}}}\frontmatter@RRAPformat}
  {}{}
}%
\makeatother
\begin{document}

\preprint{AIP/123-QED}

\title{Chip-Scale Atomic Birefringent Diffractive-Optical-Elements}
% Force line breaks with \\
\author{O. Nefesh}
 \email{ori.nefesh@mail.huji.ac.il}

\author{H. Krelman}%

\author{K. Levi}
\author{L Stern}

 \affiliation{%
Institute of Applied Physics, Hebrew University of Jerusalem, Israel%\\This line break forced% with \\
}%
\date{\today}% It is always \today, today,

\begin{abstract}
The interaction between light and vapors in the presence of magnetic fields is fundamental to many quantum technologies and applications. Recently, the ability to geometrically confine atoms into periodic structures has enabled the creation of chip‑scale, micromachined hybrid atomic‑diffractive optical elements. However, applying magnetic fields to such structures remains largely unexplored, offering potential for both fundamental and applied insights. Here, we present measurements of an atomic‑diffractive optical element subject to magnetic fields. In contrast to the well‑known polarization rotation in a Faraday medium, these diffractive atomic elements exhibit additional, rapidly oscillating rotation terms, which we validate both theoretically and experimentally. Moreover, we find that the introduction of spatially varying magnetic fields leads to a reduction in fringe visibility, which can be leveraged for gradiometric applications. Together, these effects establish a chip-scale platform where diffraction and quantum sensing are inseparably co-engineered, unveiling previously inaccessible regimes of atom–photon–magnetic interaction. By probing the magneto-optic response of periodically confined vapors, our results lay the groundwork for integrated smart-cell magnetometers and open new avenues for flat-optics-enabled quantum photonic devices. 
\end{abstract}

\maketitle

\section{\label{sec:level1}Introduction}

The interaction of atoms with magnetic fields is a pivotal area of study within the realm of light-matter physics. At the atomic level, intrinsic magnetic moments interact with external magnetic fields, which have allowed the development of various devices and technologies, including magnetometry \cite{BudkerD2007OpticalMagnetometry, Dang2010UltrahighMagnetometer}, nuclear magnetic resonance (NMR) \cite{Hall1964NUCLEARRESONANCE}, and electron spin resonance (ESR) \cite{Poole2019HandbookSpectroscopy}. Atomic magnetometers operate by detecting the Larmor precession \cite{BudkerD2007OpticalMagnetometry} and can detect scalar or vector magnetic information down to the sub-fT$/\sqrt{\mathrm{Hz}}$ level \cite{Ledbetter2008Spin-exchange-relaxation-freeVapor}. Often, the magnetic gradient is the desired measurable quantity. A common approach, coined magnetic gradiometry \cite{Sheng2017AGradiometer, Affolderbach2002AnGradiometer, Lucivero2021FemtoteslaCell, Merayo2001AGradiometer}, involves using two sensors placed at a fixed distance apart. The magnetic gradient is determined by the difference in the magnetic field between the two points divided by the spatial distance.

In recent years, the reduction of dimensions of atomic vapor systems has been extensively studied, using structures such as micromachined vapor cells\cite{Kitching2018Chip-scaleDevices}, hollow core fibers\cite{Slepkov2010SpectroscopyFibers, Russell2014Hollow-coreOptics}, ultrathin cells\cite{Sargsyan2024ElectromagneticallyCell}, evanescent interactions in prisms\cite{Matsudo1998PseudomomentumSpectroscopy, Stern2014FanoSystem} and waveguides \cite{Stern2013NanoscaleWaveguides, Zektzer2021NanoscaleReferencing}. In particular, the introduction of micromachined vapor cells has enabled the development of various chip-scale quantum sensors \cite{Kitching2018Chip-scaleDevices}, including magnetometers\cite{Schwindt2004Chip-scaleMagnetometer}, stabilized micro-frequency combs\cite{Stern2020DirectStabilization}, stabilized lasers \cite{Zhang2020UltranarrowLaser, Hummon2018PhotonicInstability}, radio-frequency \cite{Knappe2005AStability, Martinez2023AClock} and optical atomic clocks \cite{Newman2019ArchitectureClock}. Such vapor cells can be mass-produced at the wafer scale, offering high control over geometry and facilitating efficient, miniature, and cost-effective interactions between light and atoms. 

A recently demonstrated type of micromachined vapor cell is the Atomic Diffractive Optical Element (ADOE) \cite{Liron2019Chip-scaleElements}, which confines atoms to form vapor-based diffractive optical elements such as linear gratings and Fresnel lenses. Such ADOEs are realized by introducing multiple atom-filled channels by means of etching silicon wafers. The reflection or transmission of a beam from such a device can be described as a superposition of wavefront contributions emanating from the surface and the channels, after acquiring the appropriate phase. Consequently, the response of the ADOE is strongly dependent on the constituting atoms, enabling direct mapping of the atomic state to the functionality of the ADOE. Alkali vapors subjected to magnetic fields can exhibit significant induced circular birefringence, causing variations in wave properties such as polarization, amplitude, and phase. However, the exploration of birefringent atomic vapors geometrically confined within periodic gratings remains unexplored. 

Here, we investigate the magneto‑optic response of a micromachined, chip‑scale atomic‑diffractive Fresnel lens exposed to magnetic fields. First, we revisit Faraday polarization rotation and show that the lens’s intrinsic lattice birefringence adds rapidly oscillating, diffraction‑specific rotation terms, verified by a coupled atom‑diffraction model and experiment. Next, we impose a transverse magnetic‑field gradient, revealing a spatially varying phase that primarily reduces fringe visibility and can be mapped back to the local field distribution across the lens aperture. These findings clarify the magneto‑optics of geometrically confined vapors and open avenues for quantum sensing and surface‑integrated wavefront control.

\section{\label{sec:level2}Concept of birefringent ADOE}

\begin{figure*}  
    \centering
    \includegraphics[width=1\linewidth]{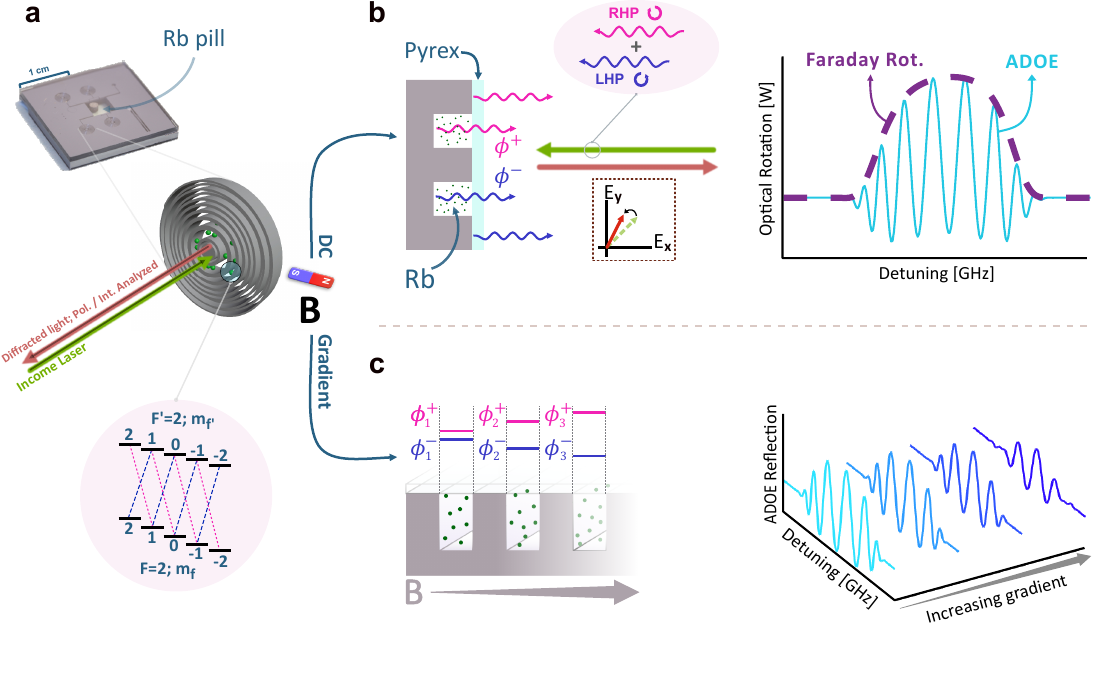}
    \caption{\textbf{Birefringence in atomic diffractive optical elements subject to an external magnetic field.d} a) Illustration of light reflecting off a Faraday-active atomic diffractive optical element subjected to a magnetic field. A linearly polarized laser beam is incident on an atomic grating, and reflected light is analyzed. The allowed atomic transitions are modified under the influence of a magnetic field, leading to interference in the reflected signal.
    b) Case of a uniform (DC) magnetic field: Linearly polarized light is separated into circular polarizations, each polarization acquiring a different phase upon interaction with rubidium. The reflected light then undergoes a polarization rotation (small box under reflected signal). On the right is an example of rapid polarization rotation, relative to conventional Faraday rotation, in the birefringent grating under such uniform fields. 
   c) Case of a spatially varying magnetic field (gradient): The relative polarization-dependent phase shift varies across the grating, leading to diminished fringe visibility in the reflected signal. On the right is an illustration of the impact of increasing the magnetic gradient on the reflected signal.}
    \label{fig: 1}
\end{figure*}

We conceptually illustrate the interaction of an atomic grating with a magnetic field in Fig. \ref{fig: 1}. We present the system’s response to a uniform magnetic field and a magnetic field gradient in Fig. \ref{fig: 1}b and Fig. \ref{fig: 1}c, respectively. To elucidate the effect of a uniform magnetic field on the Atomic Diffraction Optical Element (ADOE), we develop a model based on a one-dimensional linear atomic phase grating under the influence of a uniform magnetic field, as schematically depicted in Fig. \ref{fig: 1}b. Linearly polarized light impinging on the atomic grating will experience polarization rotation (Fig. \ref{fig: 1}a), resulting from the combination of birefringence (Faraday rotation) and the multiple interference paths introduced by diffraction. When considering a beam reflecting from the surface, one can consider the well-known thin-surface Fraunhofer diffraction approximation. This calculation effectively sums the field that is reflected from the top surface of the device and the field that is reflected after interacting with the rubidium-filled channel. In a circular polarization basis, each orthogonal polarization undergoes an atomic-state-dependent phase and amplitude response while interfering with the non-atomic portion of the reflected field as illustrated in Fig. \ref{fig: 1}b. This response, driven by the birefringent nature of rubidium, exhibits a strong dependence on the external magnetic field as a result of Zeeman splitting of the atomic sublevels (Fig. \ref{fig: 1}a). 
When considering the intensity difference between the Cartesian optical polarization components in the reflected signal from the grating, we obtain:

\begin{equation}
    I_x-I_y = I_0\cdot[\sin(\phi_+-\phi_-)+\sin(\phi_+)-\sin(\phi_-)]
    \label{eq. Pol-Rot}
\end{equation}

Where $I_0$ is the total intensity, $\phi_+$ and $\phi_-$ are the relative phases accumulated based on circular polarization, $\sigma_+$ and $\sigma_-$ (right- and left-circular polarization), respectively. This newly introduced expression (to the best of our knowledge) depends not only on the difference between the acquired phases of each polarization (such as the conventional Faraday rotation, i.e.,$ I_x-I_y  \propto \sin(\phi_+-\phi_-)$) but also on the sine of each phase. An example of rapid polarization rotation induced by a birefringent grating, in contrast to conventional Faraday rotation, is presented in Fig. \ref{fig: 1}b on the right. 
In the case of a varying magnetic field — i.e., when its amplitude changes along the grating — the relative polarization-dependent phase shift also varies along the grating (Fig. \ref{fig: 1}c). Due to variations in the accumulated phase, the reflected signal exhibits reduced fringe visibility along with a frequency offset of several fringes. A corresponding example of the resulting polarization response in the presence of a magnetic field gradient is shown in Fig. \ref{fig: 1}c on the right.
\section{\label{sec:level3}Results}
\subsection{\label{sec:level4}Constant magnetic field}
\begin{figure}
    \centering
    \includegraphics[width=0.8\linewidth]{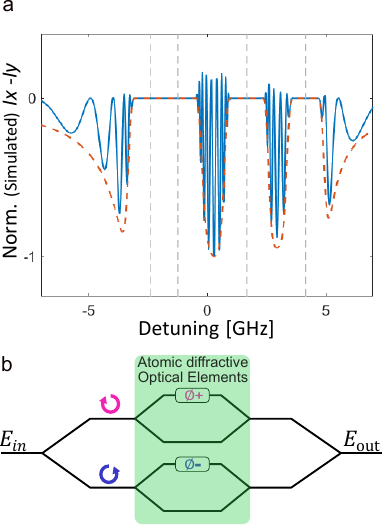}
    \caption{(a) Simulated reflection from grating VS. Faraday effect simulation of polarization rotation, $(I_x-I_y)$, in birefringence medium and atomic-dielectric grating. The orange dashed line represents the polarization rotation after reflection from a birefringence medium, corresponding to conventional Faraday rotation $I_x - I_y \propto \sin(\phi_+ - \phi_-)$, and the blue line represents the polarization after reflection from grating (1). The dashed vertical lines mark the $D_2$ rubidium absorption lines.
    (b) Mach-Zehnder analogy. Light incident on the ADOE is decomposed into circular polarization components, each of which splits into two pathways, acquiring a relative phase. The reflected signal is the coherent sum of all resulting fields.}
    \label{fig: 2}
\end{figure}

Fig. \ref{fig: 2}a presents simulations illustrating the aforementioned differences in the reflected spectra of the two configurations. The blue line represents the reflected power from an atomic-dielectric birefringent grating, while the dashed orange line corresponds to the power reflected from a uniform (i.e., nonperiodic) birefringent medium. The dashed vertical lines indicate the location of the rubidium absorption lines. In both cases, the birefringent medium is identical, consisting of Rb atoms subjected to a magnetic field. The simulations were conducted on a grating containing N=31 channels, under an external magnetic field of B=10 G and a rubidium density of $3.42 \times 10^{14}~\mathrm{cm}^{-3}$. As can be seen, reflection from a birefringent grating is dramatically different in comparison to the uniform birefringent case and exhibits rapidly oscillating fringes in between the rubidium absorption lines. This is a direct consequence of the spatially varying, circular-polarization-dependent phase profile induced by the birefringent grating. 
This process can also be thought of as a configuration of nested Mach-Zehnder (MZ) interferometers (Fig. \ref{fig: 2}b). Here, the linear polarization is split into two paths, each corresponding to circular polarization. Then, each of the circular polarizations is split into two additional paths, where a relative phase shift is introduced between the paths. Consequently, such interactions and the presence of multiple interference paths give rise to additional terms compared to conventional Faraday rotation, as shown in Eq.(\ref{eq. Pol-Rot}). 

\begin{figure}
    \centering
    \includegraphics[width=0.9\linewidth]{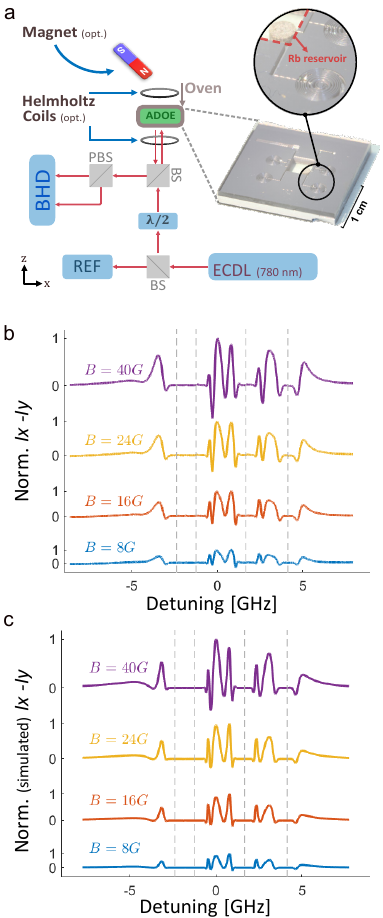}
    \caption{\textbf{Optical rotation in atomic diffractive gratings.} a) Schematics of the system. On the right, an image of the ADOE is shown with an enlarged view of the lens-shaped grating, including the rubidium reservoir. b) Experimental results showing the difference between the two Cartesian components $I_x-I_y$ corresponding to the optical rotation induced by the ADOE. c) Theoretical spectra derived from Eq.~(\ref{eq. Pol-Rot}) using the same experimental values.}
    \label{fig: 3}
\end{figure}

We now describe the experimental setup used to measure the magnetic field dependence of the interference pattern generated by the ADOE, as schematically illustrated in Fig. \ref{fig: 3}a. As previously mentioned, a Fresnel lens ADOE was selected for this study. The lens, designed to have a focal length of 7 cm, comprises of a series of circular channels etched 170 $\mu\mathrm{m}$ deep into silicon. The channel periods range from 40 $\mu\mathrm{m}$ to 120 $\mu\mathrm{m}$, with an overall diameter of 2 mm. An atomic reservoir is connected, through an additional etched channel, to the circular gratings. The ADOE was placed inside a custom-made double oven system for temperature stabilization, which was operated within a temperature range of $150^\circ$ to $180^\circ$. To induce a magnetic field, we used a pair of Helmholtz coils with a radius of 7 cm. We calculated the expected magnetic field and then measured it using a commercial Hall-probe magnetometer (Lake Shore Model 425 Gaussmeter) to ensure uniformity and accuracy. A 780 nm laser beam (corresponding to the $D_2$ line of Rb), with an approximate diameter of 1 mm and power of 20 $\mu\mathrm{W}$, irradiates the atomic lens at a normal incidence angle with respect to the lens surface. Initially, the cell is illuminated with light linearly polarized at a 45-degree angle relative to the x-axis. The reflected light was collected and then separated into Cartesian components using a polarizing beam splitter (PBS), with the intensity of each projection measured by a balanced photodetector (BPD). To understand the magnetic influence on the interference pattern and facilitate a comprehensive comparison with simulations, we conducted a series of measurements, systematically varying the induced magnetic field in the ADOE region with each iteration. The measured intensity for different cases of a constant magnetic field is presented in Fig. \ref{fig: 3}b.

By calculating the rubidium bulk susceptibility \cite{Steck2008RubidiumData} and separating it into real and imaginary parts, we extracted frequency-dependent phase and amplitude profiles for the atomic-dielectric grating. With this profile, we were able to calculate the electric field reflected from the ADOE, leading to a precise analysis of the interference pattern. By separating the calculated electric field into its circular polarizations and substituting the acquired phases into Eq.~(\ref{eq. Pol-Rot}), we obtained the expected reflected power, as shown in Fig. \ref{fig: 3}c, which was generated using the same parameters as the measurements. 
 
As mentioned, the measured and simulated intensities are shown in Figs. \ref{fig: 3}b and \ref{fig: 3}c, respectively. The flat regions in the absorption line spectra correspond to areas where the light interacts with rubidium and is fully absorbed due to the high atomic density, causing the signal to consist only of light reflected from the surface, effectively eliminating interference. In the spectral region between the absorption lines, birefringent regions emerge, where the polarization-dependent refractive index varies under the influence of the external magnetic field, resulting in the appearance of rapidly observable fringes.

\subsection{\label{sec:level5}Constant magnetic gradient}
 
We now shift our focus to examining the impact of a magnetic gradient on the interference pattern, with the vision of enabling highly compact gradiometers featuring an inherent single-output signal proportional to the magnetic gradient. As noted above, existing methods for measuring magnetic field gradients typically involve sampling the magnetic field at two spatially separated points and estimating the gradient by dividing the field difference by the distance between them. To understand the behavior of our ADOE under the influence of a magnetic gradient, we extended our previously discussed model to include a spatially varying magnetic field, which is represented as a staircase phase profile. Assuming a linear energy shift, consistent with the behavior of rubidium in the Zeeman regime, we assign a constant phase difference between two consecutive channels while averaging the phase over each channel. By adopting this approach, one can envision precise interferometric gradiometric measurements, allowing for the detection of spatial magnetic variations with high sensitivity. However, we note that adapting this approach for high-precision magnetic gradiometry would require modifications to enable access to narrow magnetic resonances. 

A known solution for the Fourier decomposition of a square wave phase modulation is given by $ c_n = \frac{-2}{\pi n} \sin\left(\phi^\pm/2\right)$   where $c_n$ is the n-th coefficient for an odd n, and $c_0=\cos(\phi^\pm/2)$. Here, $\phi^\pm$ represents the right- ($\sigma^+$) and left-circular ($\sigma^-$) polarization phase accumulation. In the presence of an external magnetic gradient, the magnetic field varies across each cell, leading to differences in the phase shift. Summing the $c_n$ coefficients for each channel in the grating yields new Fourier decomposition coefficients in the presence of a magnetic gradient. The new coefficient is given by $c_n^\nabla = A_N \cdot c_n$ for all n, where: 

\begin{equation}
    A_N = \frac{\sin\left(N \cdot \delta \phi/2\right)}{N \cdot \sin\left(\delta \phi/2\right)}
    \label{AN}
\end{equation}
\begin{figure*}
    \centering
    \includegraphics[width=1\linewidth]{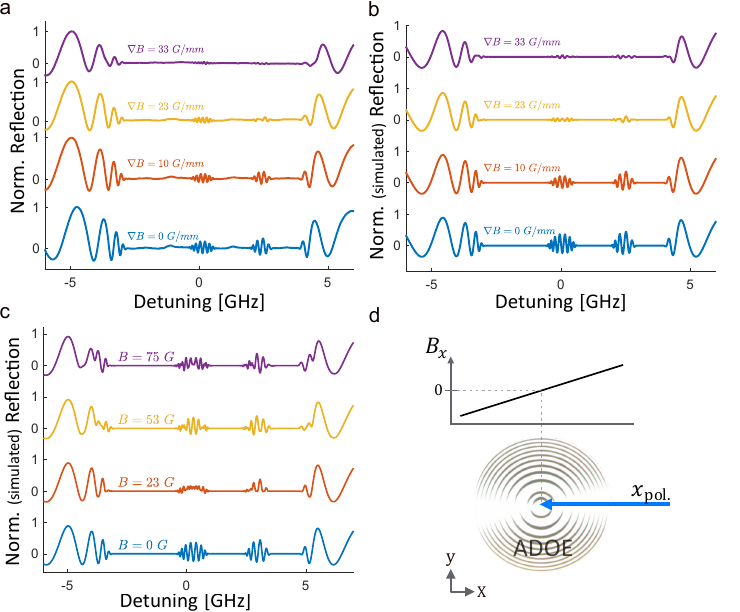}
    \caption{\textbf{Reflection spectra from an atomic Fresnel lens, subject to a permanent bar-magnet inducing a spatially varying magnetic field.} a) Measured results with a permanent magnet approaching the ADOE from the y-axis. b) Simulated spectra (magnetic gradient only). c) Simulated spectra (magnetic field only). d) Illustration of the experiment.}
    \label{fig: 4}
\end{figure*}
Here, $\delta\phi$ represents the phase difference between two consecutive channels, and N is the total number of channels in the grating. This parameter describes the loss of interference coherence caused by the phase variations across the grating. As $\delta\phi$ approaches 0, $A_N$ reduces to $sinc(N\cdot\delta\phi/2)$, reflecting the cumulative phase effect. Alternatively, using the MZ interferometer model clarifies the reduction in fringe amplitude. Due to variation in the magnetic field across the channels, each surface-channel pair in the grating functions as a Mach-Zehnder interferometer. Each interferometer contributes a slightly different phase relative to its neighbor, leading to destructive interference of light reflected after interaction with rubidium, and causing a loss of coherence of the reflected signal. After reflection, the entire signal interferes, resulting in a decrease in fringe visibility and an offset of some peaks, which originates solely from the presence of the magnetic gradient. In the case of a complex phase, the $A_N$ coefficient can account for both the rubidium absorption and interference contributions from all N channels.

We now describe the experimental setup used to measure the influence of the magnetic gradient on the interference pattern. The interference pattern was obtained using reflection spectroscopy in the presence of a spatially varying magnetic field, generated by a permanent magnet placed above the device along the y-axis, perpendicular to the table surface. A laser beam with polarization aligned along the x-direction, parallel to the table surface, was used to minimize the Voigt effect and ensure that the dominant interaction remains Faraday-like. A photodetector was placed at the end of the optical path to measure the total reflected intensity.

\begin{figure*}[t]
        \centering
    \includegraphics[width=1\linewidth]{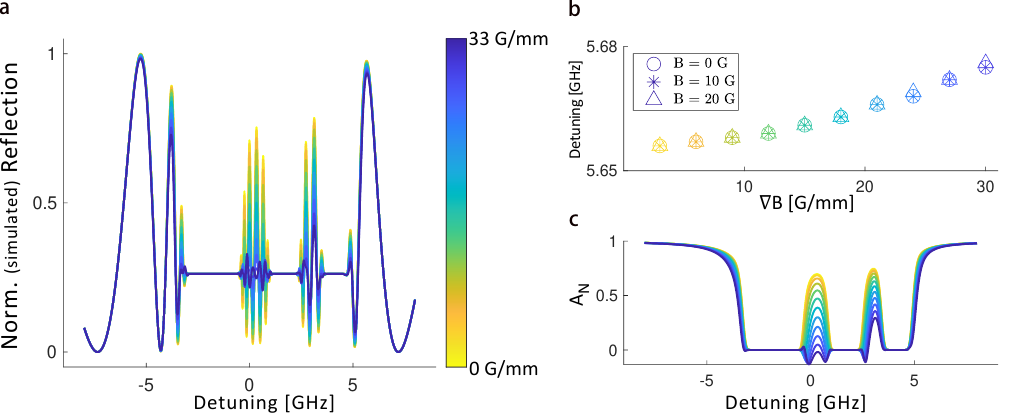}
    \caption{\textbf{Magnetic gradient effect on reflected signal.} a) Simulated reflected power under the influence of a magnetic gradient without an additional magnetic field. b) The spectral location of peak I under the influence of a magnetic gradient for various cases of external magnetic fields. c) Schematic of $A_N$, the coherence parameter, for varying magnetic gradient.}
    \label{fig: 5}
\end{figure*}

Fig.~\ref {fig: 4} illustrates the reflected power in our system under the influence of an external magnetic field (Fig.~\ref {fig: 4}a), alongside simulations based on the above-mentioned model (Figs. \ref{fig: 4}b–c), and schematics of the experiment (Fig. \ref{fig: 4}d). In Fig. \ref{fig: 4}a, we present the experimentally measured reflected power as the magnet approaches the ADOE from the y direction, showing a decrease in fringe visibility with increasing magnetic gradient. A permanent magnet generates magnetic fields in the $y$-direction, ranging from $13\,\mathrm{G}$ at a distance of $70\,\mathrm{mm}$ to $2 \times 10^3\,\mathrm{G}$ at $10\,\mathrm{mm}$, with significant accompanying gradients in the same direction, calculated and measured to range from $0.4\,\mathrm{G/mm}$ to $367\,\mathrm{G/mm}$ at these respective distances. The magnetic field in the $x$-direction vanishes along the magnet’s central $y$-axis but exhibits a strong transverse gradient, ranging from $1.7\,\mathrm{G/mm}$ to $385\,\mathrm{G/mm}$ at the same distances.
To investigate the effects of such a magnetic field, and particularly the role of the magnetic gradients on the interference pattern, we present the theoretical and experimental spectra subject to different magnetic field profiles. Since the permanent magnet generates both a magnetic field and a gradient, we simulated the reflection under two distinct conditions to assess their impact on the interference pattern: increasing the magnetic gradient (Fig. \ref{fig: 4}b) and increasing the magnetic field (Fig. \ref{fig: 4}c). A sketch of the device and its associated magnetic field configuration is shown in Fig. \ref{fig: 4}d. These conditions were aligned with the calculated and experimentally measured values induced by the permanent magnet. The magnetic field values in Fig.~\ref{fig: 4}c correspond to the maximum magnetic field at the final channel of the grating. In general, Fig. \ref{fig: 4} highlights two key features. First, one can observe the reshaping of the interference pattern. Indeed, the presence of a magnetic field shifts the rubidium atomic Zeeman levels and alters the susceptibility. The modified refractive index introduces a polarization-dependent phase profile, leading to changes in the interference pattern. For instance, this is evident in the oscillations far from the absorption lines for relatively high constant magnetic fields (the rightmost and leftmost oscillations, Fig. \ref{fig: 4}c). A second important effect is the reduction in fringe visibility, which can be attributed to the magnetic gradient causing spatially varying phase response and averaging down of the fringes, as evident in Fig. \ref{fig: 4}b.

As previously mentioned, in addition to the impact of the magnetic gradient on fringe visibility, some of the peaks also experience a spectral shift. We simulated the interference pattern to assess the effect of the magnetic gradient and gain a deeper understanding of the peaks' behavior under varying external magnetic fields. The simulations and analysis are shown in Fig. \ref{fig: 5}. In Fig. \ref{fig: 5}a, we display a simulation of the reflected power for increasing magnetic gradients, which are centered around a zero DC magnetic field. As shown, most of the peaks exhibit a decrease in magnitude, while certain peaks also display a noticeable shift in position. In this section, we analyze the behavior of the rightmost peak, labeled 'I'. In Fig. \ref{fig: 5}b, we present the position of peak 'I' as a function of the magnetic gradient, around different DC magnetic fields. Two notable observations can be made: The frequency shift shows negligible dependence on the constant magnetic field, and the derivative of the peak position shows linear behavior for small gradient values. This enables us to separate the effects of the magnetic field from the magnetic gradient, thereby improving the resolution for small gradients. In Fig. \ref{fig: 5}c, we demonstrate a visualization of the $A_N$ coefficient (Eq.~\ref{AN}), multiplied by the absorption spectra. Between the rubidium absorption lines, $A_N$ rapidly decreases with increasing gradient, reflecting the loss of coherence in the grating. On both sides, a noticeable change in the absorption spectra can be seen, resulting in the observed behavior of peak 'I'. 

\section{\label{sec:level6}Discussion and Conclusion}
To summarize, we have demonstrated the effects of both constant and spatially varying magnetic fields on an atomic diffractive grating structure. First, we derived an analytical expression for the magnetic field’s influence on light reflected from a birefringent grating, revealing new oscillatory terms in the optical polarization spectrum compared to a regular, non‐diffractive birefringent medium. These effects can be harnessed to design ultra‐thin, frequency‐sensitive devices capable of rapidly manipulating the polarization of light. Subsequently, we measured the polarization‐rotation effect induced by the birefringent grating and compared our findings with the theoretical framework, observing strong agreement. In the second part of our study, we built on our results from the constant‐field case to investigate the impact of a magnetic gradient on the interference pattern. Our analysis and experimental data show that the cumulative phase shifts introduced by the gradient produce an effective decoherence effect in the reflected signal, leading to a loss of fringe visibility accompanied by a fringe offset on either side of the absorption spectra. 
Beyond the fundamental importance of studying birefringence in periodically confined atomic systems, we suggest that these systems can exploit the effects of magnetic gradient–induced decoherence on the interference signal to simultaneously detect both the DC magnetic field and its spatial variation. As such, our device offers a compact and efficient solution for probing complex magnetic environments, with potential applications in materials science, remote quantum sensing, and magnetic field mapping.

\begin{acknowledgments}
The authors acknowledge John Kitching and Susan A. Schima for fruitful discussions and assistance with device fabrication.
\end{acknowledgments}
\section*{Data Availability Statement}
The data that support the findings of this study are available from the
corresponding author upon reasonable request.
\appendix

\section*{References}
\bibliographystyle{unsrt}
\bibliography{references_new}% Produces the bibliography via BibTeX.

\end{document}